\documentclass[12pt,preprint]{emulateapj}
\usepackage{natbib,times}
\usepackage{lscape}

\shorttitle{} \shortauthors{}
\begin{document}

\title{{Do} Ultrahigh Energy Cosmic Rays {Come from Active Galactic Nuclei and Fermi $\gamma$-ray Sources?}}

\author{Yun-Ying Jiang\altaffilmark{1,2},
L.G. Hou\altaffilmark{1}, J.L. Han\altaffilmark{1},
 X.H. Sun\altaffilmark{1},
        Wei Wang\altaffilmark{1}}

\altaffiltext{1}{National Astronomical Observatories, Chinese
Academy of Sciences,
                 20A Datun Road, Chaoyang District, Beijing 100012, China;
                 hjl@nao.cas.cn}
\altaffiltext{2}{Graduate University of the Chinese Academy of
Sciences,
                 Beijing 100049, China}


\begin{abstract}
We study possible correlations between ultrahigh energy cosmic rays
(UHECRs), observed by Auger, AGASA, and Yakutsk, and nearby active
galactic nuclei (AGNs) and $Fermi$ sources. We consider the
deflection effects by a Galactic magnetic field (GMF) model
{constrained by} the most updated measurements. We found that the
average deflection angles of UHECRs by the Galactic magnetic fields
are less than $4^\circ$. A correlation between the Auger cosmic-ray
events and nearby AGNs with a significance level of $\sim 4\sigma$
was found for the Auger UHECR data sets with or without deflection
correction. No correlation was found between the AGASA/Yakutsk
events with nearby AGNs. Marginal correlations between the Auger
events and the $Fermi$ sources, and between AGASA events and $Fermi$
AGNs were found when the deflections calculated by the GMF model
were considered. However, no correlation was found between the
Yakutsk data and $Fermi$ sources. Some $Fermi$ sources are close to
the arrival directions of UHECR events detected by Auger, AGASA, and
Yakutsk, most of which are probably chance coincidence rather than
objects producing UHECRs in the nearby Universe. Four $Fermi$
sources, NGC 4945, ESO 323-G77, NGC 6951, and Cen A, within 100~Mpc
have UHECR events within $3_.^\circ1$ from their positions, which
could potentially be cosmic-ray accelerators. However, the
association can only be confirmed if more UHECRs are preferably
detected in these directions
\end{abstract}


\section{Introduction}
The spectrum, origin, and composition of ultrahigh energy cosmic
rays (UHECRs) with energies $\geqslant$ 10$^{19}$~eV {(=10 EeV)} are
a long standing mystery in high-energy astrophysics \citep{ham84}.
\citet{gk66} and \citet{zak66} showed a theoretical distant limit
for the cosmic rays {with} energies of order 10$^{20}$~eV traveling
through the microwave background radiation field, which is called
the GZK effect. Because of the GZK effect, particles with energies
above 10 EeV are able to reach our Earth only from nearby sources
{within about} 100~Mpc. Another barrier in the investigation of the
UHECR origin is the deflections of UHECRs by the magnetic fields.
Due to the poor knowledge of the extragalactic and intergalactic
magnetic fields, the deflections of UHECRs have not yet understood.
\citet{dgst04,dgst05} suggested that the deflections by
extragalactic magnetic fields are generally less than 1$^{\circ}$,
while \citet{rdk10} and \citet{sme03} claimed that could be larger
than 10$^{\circ}$. The Galactic magnetic fields (GMFs) are
relatively better known \citep[e.g.][]{hml06,srwe08} and are widely
discussed in the studies of UHECR origin
\citep[e.g.][]{st97,tt02,ps03,nam09}. \citet{kst07} concluded that
the deflections of UHECRs by the GMFs cannot be neglected even for
the protons of $E$ = 10$^{20}$eV, since the deflection angles are
comparable with the angular resolution of current experiments.
\citet{nam09} tried {seven} GMF models to study the correlations
between UHECRs and source population(s). However, no halo component
was included in the four GMF models they used and another three GMF
models adopted from \citet{srwe08} have a strong halo component
about 7 $\mu$G. Observational constraints on the Galactic magnetic
field strength \citep{hq94,hmq99,mos96} and the configuration of
disk magnetic fields \citep{hml06,han09} should be carefully
considered in the GMF model.

Since the discovery of UHECRs \citep{lin63}, many equipments have
been used to search for these events, including Fly's Eye
\citep{fly94}, Yakutsk Extensive Air Showers Array
\citep{ikp03a,pgi05}, Akeno Giant Air Shower Array
\citep[AGASA;][]{hhik00,tsh03}, High Resolution Fly's Eye cosmic-ray
detector \citep[HiRes;][]{agasa04,hir08b} and Pierre Auger
Observatory \citep[PAO;][]{pao04,pao07,pao08}. The existence of the
GZK cutoff has been observed by the HiRes and Auger
\citep{hir08b,pao08}. Some objects have been suggested to be
possible sources of UHECRs, e.g., pulsars \citep{beo00}, active
galactic nuclei (AGNs) and subclasses of AGNs
\citep{prs92,fab98,tt01,tt01b,vbj02,gtt02,gtt04,abb06,fzb09}, radio
lobes of FR II galaxies \citep{rab93,hcf09}, and $\gamma$-ray bursts
\citep{wax95,miu05}. {However, the real sources of UHECRs are not
known yet. AGNs are favored as the most probable sources for
accelerating particles to the extreme energies \citep{ham84} for a
long time.

Recently, \citet{pao07,pao08} studied the correlation between the
arrival directions of UHECRs and the positions of nearby AGNs in the
\citet{vcv06} AGN catalog (hereafter VCV catalog). They concluded
that the arrival directions of cosmic rays with energies above
$\sim$ 60 EeV are anisotropic and UHECRs have a good correlation
with the positions of nearby AGNs ($z$~$<$~0.018). The intriguing
result attracted much attention. \citet{iaa08} found the correlation
between Yakutsk UHECRs and the nearby VCV AGNs ($\lesssim$~100~Mpc).
\citet{gfb08} investigated the correlation between the {\it Swift}
Burst Alert Telescope AGN catalog with the Auger UHECR events, and
found a correlation at a significance level of 98$\%$ when the AGNs
were weighted by their hard X-ray flux and the Auger experiment
exposure. However, some associated AGNs {of Auger events} may not
have enough energy to accelerate particles to ultrahigh energies
\citep{msp09}. The High Resolution Fly¡¯s Eye Collaboration searched
for {possible} correlation between the HiRes UHECRs and AGNs located
in the northern hemisphere; however, no significant correlation was
found.

The $Fermi$ high energy $\gamma$-ray sources are also possible UHECR
sources. The recently released $Fermi$ Large Area Telescope First
Source Catalog (1FGL) contains 1451 $\gamma$-ray point sources
\citep{fermi10} with nearly uniform sky coverage \citep{fermi09b}.
\citet{mio10} first investigated the correlation {between} Auger
UHECRs and 1FGL sources without considering the deflection by the
GMFs and redshifts of $Fermi$ AGNs, and concluded that the UHECRs
are not associated with $Fermi$ sources. The possible correlation of
UHECRs and $Fermi$ sources should be re-examined after the GZK
cutoff and the UHECR deflection by the GMFs are considered.

In this work, we construct a new GMF model based on the updated
measurements of the Galactic magnetic fields and investigate the
deflections of UHECRs by the GMFs. Considering the GZK cutoff and
the deflection correction through our GMF model, we re-examine the
possible correlation between UHECRs and nearby AGNs and $Fermi$
sources. In Section 2, we discuss available data of UHECRs detected
by Auger, AGASA, and Yakutsk and possible astrophysical objects. The
deflections of UHECRs by the GMFs are discussed in Section 3. The
correlation studies are given in Section 4. Discussions and
conclusions are presented in Section 5.

\section{UHECR Data and Potential Cosmic ray sources}

\subsection{UHECR events}

High-quality UHECR data and reasonable deflection correction by the
GMF models are crucial to understand the origin of UHECRs.
Therefore, we only consider UHECR events which satisfy two criteria:
(1) good angular and energy resolutions; (2) ultrahigh energy
($E>40$~EeV) which has a predictable small deflection angle.

We work on the UHECR events recorded by Auger, AGASA, and Yakutsk.
The HiRes UHECRs have a typical angular resolution of $0_. ^\circ6$
\citep{agasa04} and 27 events have been published. However, we did
not use them since the detail positions and energies are not
available \citep[][and references therein]{kas06}.

Auger is located in Argentina and began to collect data from 2004
January 1. It has two systems, one to measure fluorescence in the
atmosphere and the other to detect Cerenkov light from relativistic
particles. The angular resolution of Auger is about $0_.^\circ9$
\citep{ave07}. Eighty-one events with $E>40$ EeV {have been recorded
by Auger} from 2004 January 1 to 2007 August 31 but only 27 events
with energies above 57 EeV were published. Recently, another 31 new
events ($E>57$ EeV) were detected \citep{pao09}; however, data are
not yet available. AGASA and Yakutsk are all located in the northern
hemisphere. AGASA has been operated for 12 years, and ceased
operation on 2004 January 4 Its angular resolution is about
$1_.^\circ8$ \citep{hhik00}. Up to 2000, 57 events with $E>40$~EeV
have been published \citep{hhik00}. Yakutsk collected 51 events with
$E>40$~EeV, its angular resolution is smaller than 5$^\circ$
\citep{pgi05}. Therefore, the UHECRs data used in this work include
135 events, 57 recorded by AGASA and 51 by Yakutsk in the northern
hemisphere with $E>40$ EeV, and 27 {events recorded} by Auger in the
southern hemisphere with $E\geqslant57$~EeV.

Due to the different angular resolution, different energy
calibration, and different sky exposure for Auger, AGASA, and
Yakutsk UHECR events, we search for the possible correlations
separately between the three sets of UHECRs and astrophysical
objects.

\subsection{Potential Astrophysical Objects as Cosmic-Ray Sources}

The nearby AGNs used in this work are extracted from the 13{th} VCV
AGN catalog \citep{vcv10}, with a redshift limit of
$z\leqslant0.024$ which corresponds to a GZK cutoff $\sim100$ Mpc
for the Hubble constant $H_0=71\rm~km~s^{-1}~Mpc^{-1}$. The VCV
catalog includes all known AGNs reported in the literature. There
are 133,336 quasars, 1374 BL Lac objects, and 34,231 active
galaxies. This catalog is not complete and not uniform due to
different selection criteria and telescope time devoted to different
sky areas. On the other hand, the difficulties in the classification
of galaxies with weak AGN-like activity result in the confusion in
identification of such sources. Thus, there are some flaws when
using some given particular AGN catalog to study their possible
correlations with UHECRs \citep[see ][]{msp09}. AGNs are favored as
the most probable sources for accelerating particles to the extreme
energies \citep{ham84} and almost all known AGNs are listed in the
catalog. In this work, we work on the VCV nearby AGNs and compare
the results with that by \citet{pao07}. For the VCV AGN catalog, the
incompleteness is particularly serious around the galactic plane of
$|b|~\leqslant~10^\circ$. Therefore, only the 830 nearby AGNs with
$|b|>10^\circ$ are used in this work.

Another type of UHECR source candidates investigated in this work
are $Fermi$ sources. \citet{fermi09} presented the initial 3 month
results for the 205 most significant $\gamma$-ray sources with
energies above 100 MeV. Recently, after one-year observations,
Fermi/LAT released the catalog for the all-sky 1451 $\gamma$-ray
sources \citep{fermi10}, which contains 820 identified sources and
631 unidentified sources. The Fermi/LAT first-year $\gamma$-ray
source catalog (1FGL) is a complete $\gamma$-ray source sample with
a $\gamma$-ray flux threshold of $>$ 4$\times$10$^{-10}$ cm$^{-2}$
s$^{-1}$ in the energy range 1 -- 100 GeV. \citet{fermi10}
identified many $Fermi$ AGNs with redshift data. Due to the GZK
cutoff, we restrict the $Fermi$ sources with $z\leqslant0.024$.
There are only eight AGNs that satisfy this criterion. In this work,
we adopt these eight objects and other $Fermi$ sources without
redshift data to search for their possible correlation with UHECRs.
Because the diffuse emission dominates at low Galactic latitude
($|b|~\leqslant~10^{\circ}$), the $Fermi$ sources of
$|b|~\leqslant~10^{\circ}$ are also discarded. In total, 635 $Fermi$
sources are used in this work, including 262 identified sources (235
AGNs) and 373 unidentified sources.

\section{The Galactic Magnetic Fields and Deflection of UHECRs}

The GMFs have large-scale regular and small-scale turbulent
components. The deflection angles of UHECRs caused by the turbulent
fields are typically 1 order of magnitude smaller than that by the
regular fields \citep{tt05}; therefore, we ignore the turbulent
component in this work. The Galactic magnetic fields in general are
described as the regular magnetic field in the disk and the possible
large-scale field in the halo. Previous magnetic field models for
the disk are either axissymmetric \citep[e.g.][]{st97} or
bi-symmetric \citep[e.g.][]{hmr99,tt02}. However, none of these
simple models agrees well with the observations
\citep{mfh08,srwe08}.

\begin{figure}[bth]
\centering
\includegraphics[angle=-90,width=8cm]{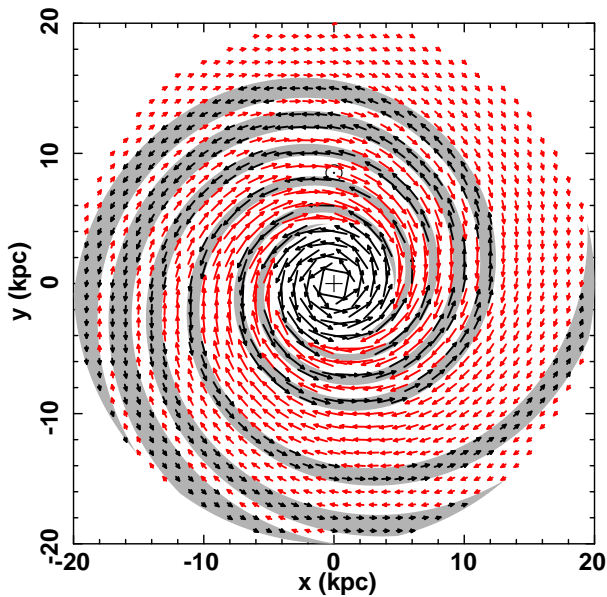}
\caption{Configuration of the disk magnetic field. The shaded area
outlines spiral arms given by \citet{hhs09}.} \label{df}
\end{figure}

Here we developed a toy model based on rotation measures (RMs) of
pulsars by \citet{hml06}. Magnetic fields in the disk are reversed
from arms to inter-arms which has recently been verified by
\citet{nk10}. The radial profile of the field strength can be
described as $B(R)=B_0\exp(-(R-R_0)/R_B)$, where $B_0=2.1$~$\mu$G is
the local field strength, $R$ is the distance from the Galactic
center, $R_0=8.5$~kpc is the galactocentric distance of the Sun, and
 the scale radius is $R_B=8.5$~kpc. Here, we
use the four-arm model of \citet{hhs09} to describe the spiral
structure of our Milky Way. In the polar coordinates, the $\it i$th
arm can be described as $R=R_i\exp[(\theta-\theta_i)\tan\psi_i]$,
where $R_i$ is the initial radius, $\theta_i$ is the start azimuthal
angle, and $\psi_i$ is the pitch angle of the arm. The values of
these parameters of four arms are given in Table 1 of \citet{hhs09}.
To ease the model description, magnetic fields within 4.6 kpc do not
have reversals. The initial width of each arm is set to be 0.4 kpc
in our work. The pitch angle of the magnetic field is $-11^\circ$ as
used by \citet{hml06}. The configuration of the disk magnetic fields
in our toy model is displayed in Figure~\ref{df}, which gives the
counterclockwise field in the arms and the clockwise field in the
inter-arm regions. Here, we remind that this is only a toy model and
the four-arm model is not the best one to match all tracers for the
spiral arms \citep{hhs09}.

The halo magnetic fields consist of a dipole poloidal field and a
toroidal field with opposite directions above and below the Galactic
plane. The field configuration was derived from the antisymmetric RM
sky revealed by the extragalactic radio sources \citep{hmb97,hmq99}
and the vertical filaments in the Galactic center
\citep{ymc84,ymc04}. At present, it is not clear whether such
toroidal fields extend from the solar vicinity to the Galactic
center and what the scaleheight they have. Here we use the formula
given by \citet{ps03} in the Cartesian coordinates to describe the
toy model:
\begin{equation}
\left\{
\begin{array}{ll}
B_x=-B_T \sin(\phi)\rm sign(z)\\
 B_y=B_T \cos(\phi)\rm sign(z),
 \end{array}
 \right.
\end{equation}

where $B_T$ is given by

\begin{equation}
B_T=
 \left\{
\begin{array}{ll}
B_{\max}\frac{1}{1+(\frac{z-H}{P})^2}, & \sqrt{x^2+y^2}<R \\
B_{\max}\frac{1}{1+(\frac{z-H}{P})^2}\exp(-\frac{(x^2+y^2)^{1/2}}{R}),
& \sqrt{x^2+y^2}>R
\end{array}
\right.
\end{equation}
Here $R=15$ kpc is the radius of the toroidal field, $H=1.5$ kpc is
the scaleheight of the toroidal disk, $P$~=~0.3~kpc is the halfwidth
of the Lorentzian distribution, and the maximal value of the
magnetic field is $B_{\max}=1$ $\mu$G.

The poloidal field is modeled as a dipole with a cylinder (height
300 pc, diameter 200 pc) in the Galactic center. The existence of
the dipole field is questioned by \citet{tss09} and \citet{mgh10}.
However, their data in the Galactic pole region clearly support the
local vertical field. In this work, we try to include the dipole
field in our toy model. In the polar coordinates the poloidal field
strength is (in the $xz-$plane, $\theta$ ranges from 0 to $\pi$ and
goes from north to south pole)

\begin{equation}
B=B_P\sqrt{3\cos^2(\theta)+1}.
\end{equation}
Then the Cartesian components of the poloidal field are

\begin{equation}
\left\{
\begin{array}{rrl}
B_x &=& 1.5B_P\sin 2\theta \cos \phi\\
B_y &=& 1.5B_P\sin 2\theta \sin \phi\\
B_z &=& B_P(3 \cos^2 \theta - 1),
\end{array}
\right.
\end{equation}
where
\begin{equation}
B_P= \left\{
\begin{array}{lllllll}
100/R_P^3   & & & 5 < R_P < 15\\
1 & & & 2<R_P<5\\
0.2/R_P^3& & & 0.1 < R_P < 2\\
2000 & & & R_P < 0.1 ~and~ |z|<0.15.
\end{array}
\right.
\end{equation}
Here $R_P=\sqrt{x^2+z^2}$. $R_P$ and $z$ {are} in units of kpc, and
$B_P$ is in units of $\mu$G. The constants in $B_P$ were selected to
meet the characters of observed filaments in the Galactic center
\citep[1-2 $m$G;][]{mos96}, and a 0.2 $\mu$G vertical field
component in the vicinity of the Sun \citep{hq94}.

In our toy model, the field transition from the arms to the
inter-arms is not smooth, the influence of the bar in the Galactic
center is not considered yet. The detailed model is beyond the scope
of this paper. We will investigate how much UHECRs are reflected by
each magnetic field component in the model.

\begin{figure*}
\includegraphics[angle=0,width=0.48\textwidth]{arm_disk.eps}
\includegraphics[angle=0,width=0.48\textwidth]{arm_dp.eps}
\includegraphics[angle=0,width=0.48\textwidth]{arm_dipole.eps}
\includegraphics[angle=0,width=0.48\textwidth]{arm_dt.eps}
\includegraphics[angle=0,width=0.48\textwidth]{arm_toroidal.eps}
\includegraphics[angle=0,width=0.48\textwidth]{arm_pt.eps}
\includegraphics[width=0.48\textwidth]{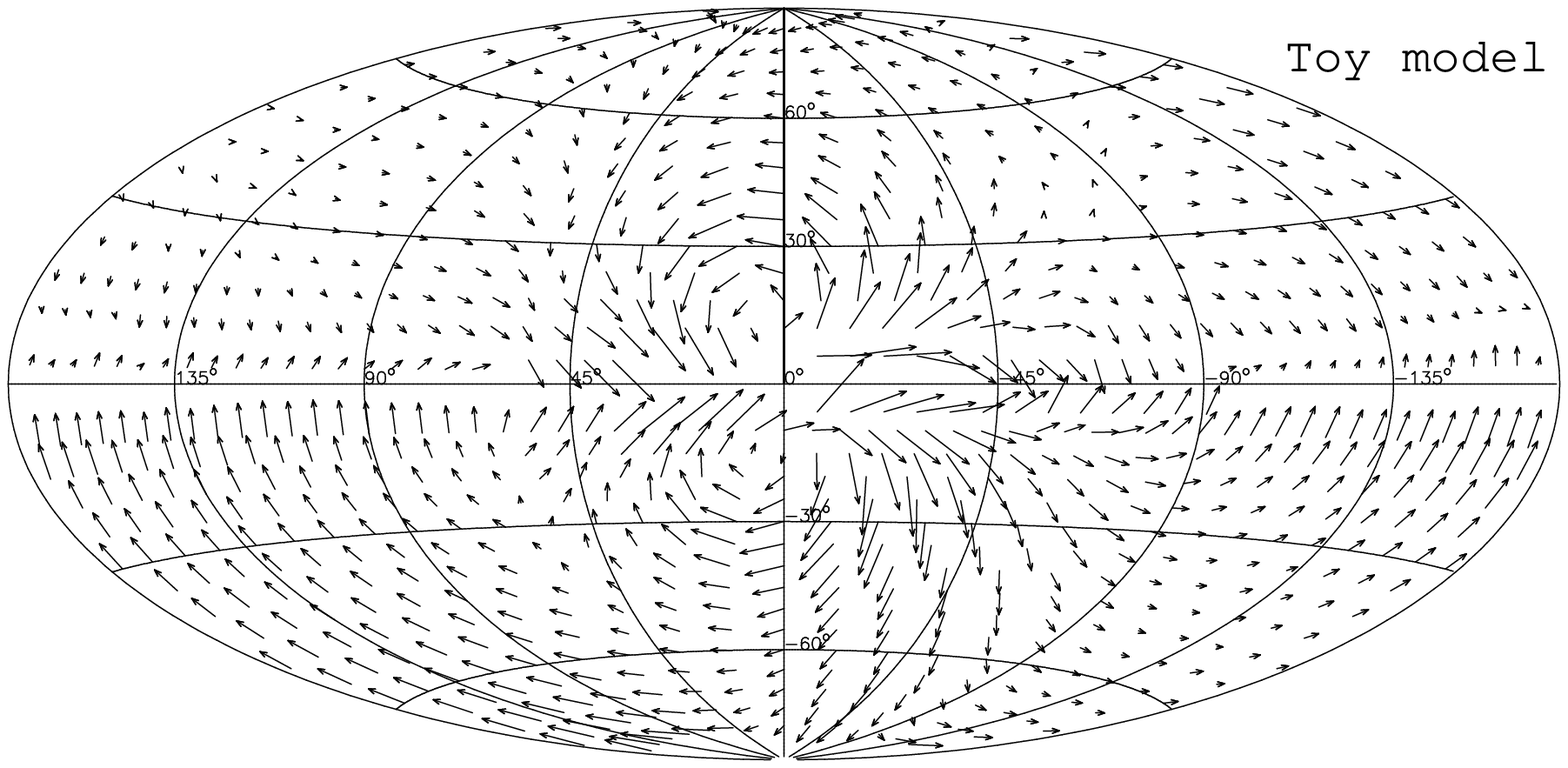} \centering
\includegraphics[angle=0,width=0.48\textwidth]{arm_ps.eps}
\caption{Deflection maps of CR protons of various {GMF components
and their} combinations in our toy model and the PS model
\citep{ps03}. The energy of CRs is fixed to be 40 EeV. All the maps
are plotted in the Galactic coordinates. Each arrow goes from the
observed direction of a CR on the Earth to the actual incoming
direction of the source.} \label{fig2}
\end{figure*}

\begin{figure*}
\includegraphics[angle=-90,width=130mm]{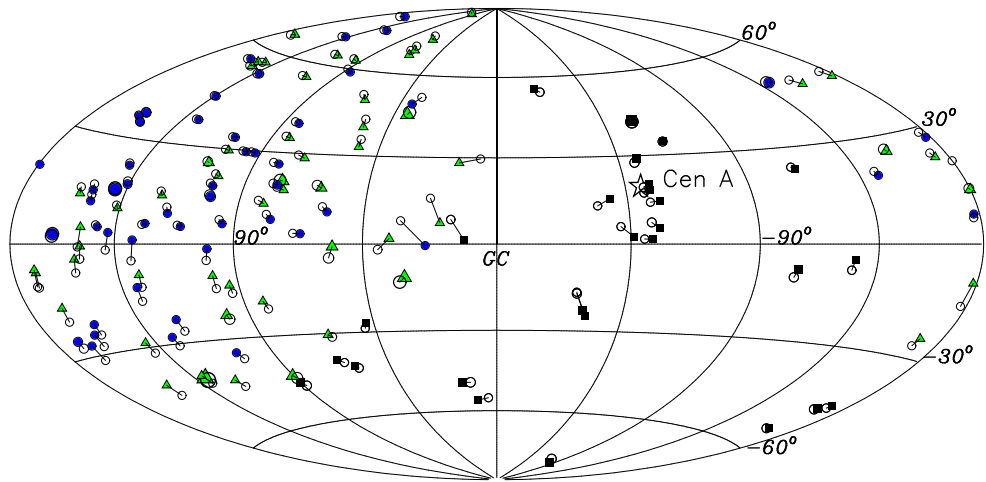} \centering
 \caption{Deflection of 135 {detected} UHECRs according to
 our GMF {toy} model (in the Galactic
   coordinates). The circles are the original arrival directions
   of 135 UHECRs observed on the Earth, the squares, triangles, and dots are the deflection
    corrected positions of the Auger, AGASA, and Yakutsk CR events, respectively.
    The size of symbols is proportional to the CR energy. {The star marks the Cen A.}}
\label{fig3}
\end{figure*}

\begin{figure}
\centering
\includegraphics[angle=-90,width=70mm]{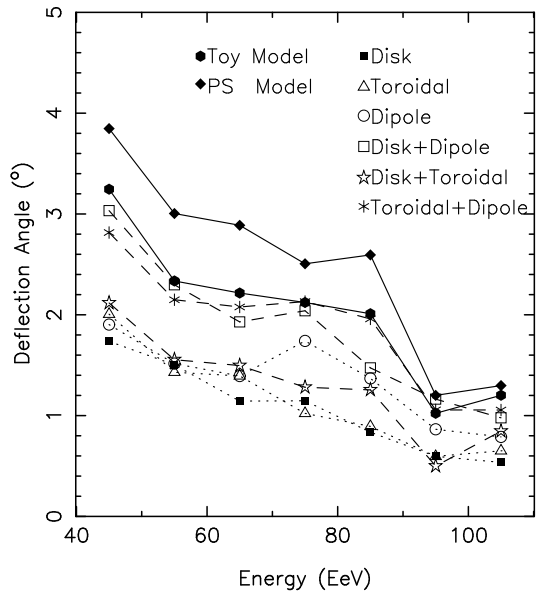}
 \caption{Average deflection angles of 135 {detected} UHECRs as a function of energy
 by {our GMF toy} model and the PS model \citep{ps03}. The solid lines are our model (black dot) and PS model (black diamond),
 dotted lines are three GMF components, dashed lines are their combinations.}
\label{fig9}
\end{figure}

The cosmic rays are deflected in the GMFs because of the Lorentz
force. Following \citet{kst07}, the net deflection can be
approximated as

\begin{equation}
{\mathbf \Theta}\approx \frac{Z\,q_e}{pc}\int d\vec{l} \times \vec
B_t\,
\end{equation}

where $Zq_e$~is the charge of cosmic-ray particles, $p$ is the
momentum along the line of sight (LOS), and  $\vec B_t$ is the field
component perpendicular to the LOS. The integral is along the LOS
from the source to the observer. In our work, the Hammurabi code
\citep{wjrk09} is used to calculate the all-sky deflections.

In Figure~\ref{fig2}, we show the deflection angle maps of UHECRs
for various magnetic field components and their combinations in the
toy model, and also the PS model \citep{ps03} which has a different
disk component. The energy of CR protons is fixed to be 40 EeV. The
disk field generates a strong deflection near the Galactic disk and
in the high latitude. The toroidal field is similar to the disk
field and has a large deflection near the disk, while the
orientations are opposite above and below the disk. The dipole
component has very strong deflections in the Galactic center region.
Deflection maps by our toy model and the PS model  have some similar
features: (1) the maximal deflections take place in the inner
Galactic disk and Galactic central regions; (2) in the southern sky
of the outer Galaxy, the deflection is as strong as that in the
Galactic center, while in the northern sky, a very small deflection
occurs. This is a result of combining the toroidal and the dipole
magnetic field components. However, there is quite difference in
some regions between these two models: in the northern sky, our
model generates smaller deflections from $l=0^\circ$ to
$l=180^\circ$, and the orientation is contrary to the PS model in
the region of $-90^\circ<l<-180^\circ$; in the southern sky, the PS
model has larger deflections, particularly in the region of
$-90^\circ<l<-180^\circ$.

The energy of observed UHECRs in our data sets varies from 40 EeV to
more than 200 EeV. The deflections for the detected UHECR events
according to our toy model are shown in Figure~\ref{fig3}.
Figure~\ref{fig9} shows the average deflection angles of 135 UHECRs
by the two GMF models and the various magnetic component
combinations. We found that the deflections decrease with the CR
energy as expected. Magnetic fields in our model generate smaller
deflections than those in the PS model. CR events with energy below
90 EeV are deflected by an angle $\lesssim 3^\circ$. The arrival
directions of the events with energy above 90 EeV are deflected less
than 2$^\circ$. The deflection angles are generally similar to the
angular resolution of the UHECR detectors ( $\sim 1^\circ
-2^\circ$); therefore the GMF deflection correction is still desired
in understanding the origin of UHECRs.


\section{Correlations Between the UHECR Events and Potential Cosmic-Ray Sources}

In this work, we try to correct the deflections of the UHECR events
by the GMFs, and then search for possible correlations between three
UHECR data sets (Auger, AGASA, and Yakutsk) and nearby AGNs and
Fermi sources. We will compare the correlation results with and
without the deflection correction. Here, we first introduce the
correlation analysis method.

\begin{table*}
\centering
 \caption[]{Correlation {Between UHECRs with AGNs and Fermi $\gamma$-Ray Sources With or Without Deflection Corrections by GMFs} }
 \label{tab4}
 \setlength{\tabcolsep}{0.1mm}
 \small
 \begin{tabular}{clccccccccccccccccccccc}
  \hline\noalign{\smallskip}
    \hline\noalign{\smallskip}
  & & & \multicolumn{3}{l}{Auger Events}&  & &&& &\multicolumn{5}{l}{AGASA  Events}&  && & &\multicolumn{3}{l}{Yakutsk  Events} \\
  \hline\noalign{\smallskip}
     (1)& (2) &(3) &(4) & (5)& (6)  &(7)&& &&(8)& (9) &(10) & (11) &(12) & &&&(13) &(14) &(15) & (16) &(17)\\
 Sources & GMF&$\delta$~($^\circ$)&$N_{cs}$&$\overline{N}_{mc}$& $P$ & $\sigma$
 && &&$\delta$~($^\circ$)&$N_{cs}$& $\overline{N}_{mc}$& $P$ &$\sigma$& &&&$\delta$~($^\circ$)&$N_{cs}$&$\overline{N}_{mc}$& $P$ &$\sigma$  \\
  \hline\noalign{\smallskip}
VCV/AGN &   Non   &   2.4     &   17  &   6.7     &   1$\times10^{-6}$    &   4.7    &   &   &   &   1.4     &   11  &   6.7     &   0.066   &   1.8    &   &   &   &   3.0     &   19  &   17.9    &   0.420   &   0.4    \\
    &   Toy model&   3.6     &   20  &   10.5    &   2$\times10^{-5}$    &   4.1    &   &   &   &   2.1     &   17  &   11.5    &   0.049   &   1.9    &   &   &   &   3.6     &   25  &   21.4    &   0.155   &   1.2    \\
    &   Disk    &   3.3     &   21  &   9.5     &   $<$1$\times10^{-6}$   &   5.0    &   &   &   &   0.6     &   4   &   1.4     &   0.048   &   2.3    &   &   &   &   0.3     &   1   &   0.3     &   0.277   &   1.2    \\
    &   Dipole  &   2.1     &   14  &   5.5     &   1.6$\times10^{-5}$  &   4.2    &   &   &   &   2.7     &   22  &   18.3    &   0.172   &   1.1    &   &   &   &   3.3     &   22  &   19.9    &   0.307   &   0.7    \\
    &   Toroidal    &   3.3     &   20  &   9.5     &   $<$1$\times10^{-6}$   &   4.6    &   &   &   &   3.6     &   29  &   25.1    &   0.164   &   1.1    &   &   &   &   3.0     &   20  &   17.4    &   0.237   &   0.9    \\
    &   Disk+Dip.  &   3.0     &   18  &   8.5     &   2$\times10^{-5}$    &   4.2    &   &   &   &   1.5     &   13  &   7.1     &   0.019   &   2.4    &   &   &   &   2.4     &   15  &   13.4    &   0.350   &   0.5    \\
    &   Disk+Toro.  &   3.9     &   21  &   11.4    &   2$\times10^{-5}$    &   4.2    &   &   &   &   2.4     &   18  &   14.6    &   0.179   &   1.1    &   &   &   &   3.0     &   21  &   17.4    &   0.148   &   1.2    \\
    &   Toro.+Dip.   &   3.3     &   19  &   9.5     &   2$\times10^{-5}$    &   4.2    &   &   &   &   2.4     &   15  &   14.6    &   0.499   &   0.1    &   &   &   &   3.6     &   25  &   21.4    &   0.155   &   1.2    \\
    &   PS model  &   3.0     &   18  &   8.5     &   2$\times10^{-5}$    &   4.2    &   &   &   &   1.5     &   11  &   6.7     &   0.062   &   1.8    &   &   &   &   1.2    &   7   &   4.2     &   0.117   &   1.5   \\
1FGL/All    &  Non    &   4.0     &   17  &   13.5    &   0.091   &   1.5    &   &   &   &   3.0     &   25  &   20.9    &   0.153   &   1.2    &   &   &   &   2.6     &   16  &   11.7    &   0.096   &   1.5    \\
    &   Toy model &   0.9     &   5   &   1.0     &  { 2.5}$\times10^{-3}$  &   4.1    &   &   &   &   2.4     &   18  &   12.9    &   0.070   &   1.7    &   &   &   &   4.2     &   28  &   22.7    &   0.053   &   1.8    \\
    &   Disk    &   2.4     &   10  &   6.1     &   0.056   &   1.9    &   &   &   &   2.7     &   22  &   17.1    &   0.095   &   1.5    &   &   &   &   4.2     &   29  &   23.9    &   0.063   &   1.7    \\
    &   Dipole  &   4.2     &   19  &   14.3    &   0.024   &   2.1    &   &   &   &   5.7     &   45  &   42.0    &   0.180   &   1.1    &   &   &   &   4.2     &   29  &   23.3    &   0.042   &   1.9    \\
    &   Toroidal    &   0.6     &   2   &   0.4     &   0.073   &   2.4    &   &   &   &   4.8     &   41  &   35.2    &   0.038   &   1.9    &   &   &   &   5.1     &   32  &   27.7    &   0.080   &   1.6    \\
    &   Disk+Dip. &   1.8     &   9   &   3.7     &   5.3$\times10^{-3}$  &   3.0    &   &   &   &   1.8     &   14  &   8.4     &   0.033   &   2.1    &   &   &   &   3.6     &   25  &   19.7    &   0.066   &   1.7    \\
    &   Disk+Toro. &   1.8     &   7   &   3.7     &   0.064   &   1.9    &   &   &   &   3.6     &   31  &   24.9    &   0.056   &   1.7    &   &   &   &   5.1     &   33  &   27.7    &   0.034   &   2.0    \\
    &   Toro.+Dip.  &   3.6     &   17  &   11.3    &   9.1$\times10^{-3}$  &   2.5    &   &   &   &   5.4     &   42  &   38.0    &   0.106   &   1.4    &   &   &   &   3.9     &   27  &   20.7    &   0.028   &   2.1    \\
    &   PS model  &   2.1     &   12  &   4.9     &   6.0$\times10^{-4}$   &   3.7    &   &   &   &   1.5     &   11  &   5.6     &   0.019   &   2.4    &   &   &   &   3.3     &   23  &   16.9    &   0.042   &   1.9    \\
1FGL/AGN    &   Non   &   7.6     &   20  &   14.7    &   0.013   &   2.4    &   &   &   &   3.8     &   23  &   15.8    &   0.025   &   2.2    &   &   &   &   5.2     &   21  &   19.8    &   0.416   &   0.4    \\
    &   Toy model&   0.9     &   2   &   0.3     &   0.037   &   3.1    &   &   &   &   2.7     &   15  &   7.8     &   7.0$\times10^{-3}$  &   2.8    &   &   &   &   3.6     &   15  &   11.6    &   0.150   &   1.2    \\
    &   Disk    &   8.1     &   20  &   15.1    &   0.010   &   2.4    &   &   &   &   2.7     &   15  &   8.6     &   0.020   &   2.4    &   &   &   &   0.6     &   1   &   0.4     &   0.343   &   0.9    \\
    &   Dipole  &   3.3     &   9   &   4.0     &   9.3$\times10^{-3}$  &   2.8    &   &   &   &   4.5     &   27  &   20.0    &   0.032   &   2.0    &   &   &   &   4.2     &   17  &   14.9    &   0.293   &   0.7    \\
    &   Toroidal    &   7.5     &   19  &   13.8    &   0.012   &   2.4    &   &   &   &   4.2     &   20  &   17.4    &   0.255   &   0.8    &   &   &   &   3.6     &   12  &   11.6    &   0.498   &   0.2    \\
    &   Disk+Dip. &   1.8     &   4   &   1.2     &   0.030   &   2.7    &   &   &   &   1.8     &   10  &   4.0     &   5.2$\times10^{-3}$  &   3.2    &   &   &   &   3.6     &   16  &   12.2    &   0.124   &   1.3    \\
    &   Disk+Toro. &   8.1     &   20  &   15.1    &   0.010   &   2.4    &   &   &   &   3.6     &   20  &   13.4    &   0.027   &   2.1    &   &   &   &   3.9     &   15  &   13.1    &   0.305   &   0.7    \\
    &   Toro.+Dip. &   3.9     &   9   &   5.1     &   0.045   &   2.0    &   &   &   &   1.8     &   6   &   3.9     &   0.183   &   1.1    &   &   &   &   3.9     &   15  &   13.1    &   0.305   &   0.7    \\
    &   PS model &   5.4     &   13  &   8.7     &   0.047   &   1.9    &   &   &   &   1.5     &   9   &   2.6     &   7.0$\times10^{-4}$  &   4.1    &   &   &   &   3.3     &   12  &   10.3    &   0.312   &   0.6    \\
1FGL/Un-id  &   Non   &   7.0     &   21  &   18.3    &   0.100   &   1.5    &   &   &   &   6.0     &   36  &   32.6    &   0.201   &   1.0    &   &   &   &   2.6     &   10  &   5.5     &   0.038   &   2.1   \\
    &   Toy model&   3.0     &   11  &   6.1     &   0.021   &   2.4    &   &   &   &   5.4     &   29  &   25.6    &   0.196   &   1.0    &   &   &   &   6.6     &   30  &   23.3    &   0.015   &   2.2    \\
    &   Disk    &   6.0     &   19  &   15.3    &   0.051   &   1.8    &   &   &   &   5.4     &   32  &   27.8    &   0.151   &   1.2    &   &   &   &   5.1     &   23  &   17.4    &   0.052   &   1.8    \\
    &   Dipole  &   1.2     &   4   &   1.2     &   0.031   &   2.6    &   &   &   &   6.0     &   37  &   31.9    &   0.093   &   1.5    &   &   &   &   6.9     &   30  &   25.2    &   0.072   &   1.6    \\
    &   Toroidal    &   8.1     &   21  &   19.0    &   0.120   &   1.5    &   &   &   &   6.3     &   35  &   32.2    &   0.251   &   0.8    &   &   &   &   5.1     &   23  &   16.6    &   0.029   &   2.1    \\
    &   Disk+Dip.  &   0.9     &   3   &   0.7     &   0.030   &   2.9    &   &   &   &   6.3     &   38  &   32.2    &   0.052   &   1.7    &   &   &   &   6.0     &   28  &   21.6    &   0.029   &   2.0    \\
    &   Disk+Toro. &   7.8     &   21  &   18.6    &   0.085   &   1.6    &   &   &   &   6.0     &   32  &   30.0    &   0.335   &   0.6    &   &   &   &   5.7     &   28  &   19.3    &   2.9$\times10^{-3}$   &   2.8    \\
    &   Toro.+Dip.  &   3.6     &   12  &   8.1     &   0.066   &   1.7    &   &   &   &   5.1     &   29  &   24.9    &   0.151   &   1.2    &   &   &   &   7.5     &   32  &   26.8    &   0.041   &   1.9    \\
    &   PS model &   1.8     &   8   &   2.5     &   1.7$\times10^{-3}$   &   3.7    &   &   &   &   0.6     &   1   &   0.5     &   0.387   &   0.7    &   &   &   &   6.6     &   30  &   23.9    &   0.028   &   2.0    \\

 \noalign{\smallskip}\hline

\end{tabular}
\end{table*}

\subsection{The Correlation Analysis Method}

To carry out the correlation analysis, we use the angular
correlation function method described in
  \citet{tt01,tt01b}, \citet{gtt02,gtt04}, and \citet{gt05}. For a
  sample with $n_r$ cosmic-ray events, we count the number $N_{cs}$
  for UHECR-source pairs within a given angle $\delta$, which is
  called ``bin size'' and various from 0 to a large angle. We count 1 if
  at least one potential cosmic-ray source (such as AGNs) falls into ``the
  bin'' and count 0 if no source falls into the bin. To check the
  chance probability, we use {Monte Carlo} simulations. We generate
  a large number (e.g., with $N$= 10$^4$, 10$^6$ for some cases)
  of simulated sets of UHECR events, each set has the
  same number of events as the real sample. The simulated UHECRs are
  isotropic and the locations of simulated events are random. The
  distribution of locations is constrained by the overall
  exposure of the UHECR detectors. For a given $\delta$, we first count the number of UHECR-source
  pairs  $N_{mc}$ for each simulated UHECR sample, and obtain
  a mean $\overline{N}_{mc}$ and the variance
  $\sigma_{mc}$ from $N$ simulation sets. The exposure of Auger, AGASA, and Yakutsk
   depends on the celestial declination \citep{pao08, thh99, ikp03}. The Auger exposure
  used in our work is obtained from the fitting of the declination
  distribution of low-energy Auger events ($E$~$<$~10~EeV, from the Auger
  Web site\footnote{http://auger.colostate.edu/ED/}) with a third-order
  polynomial function. The AGASA
  exposure function is taken from \citet{thh99}. The Yakutsk exposure function is derived from \citet{ikp03}. The probability that the
  observed UHECR-source pairs are in coincidence {with} a
  random distribution is estimated by

\begin{equation}
 P(\delta)=\frac{{\rm number~of~simulated~sets~with~} N_{mc}\geqslant
 N_{cs}}{N}.
\end{equation}

The significance of the correlation can be defined as
$\sigma(\delta)=\frac{N_{cs}-\overline{N}_{mc}}{\sigma_{mc}}$. We
emphasize that $P$ and $\sigma$ vary with $\delta$. Larger chance
probability $P(\delta)$ indicates that the observed pairs of cosmic
rays and astrophysical objects are more likely the statistical
coincidence of random isotropic UHECR events. The higher
significance corresponds to a smaller $P(\delta)$ which suggests
that the objects of the pairs are more likely to be potential UHECR
sources. In this paper, $P<10^{-2}$ is believed as an indicator of
some correlation. We use this method to evaluate the possible
correlations between the Auger/AGASA/Yakutsk events and the possible
cosmic-ray sources.

\begin{figure*}[bth]
\includegraphics[width=130mm]{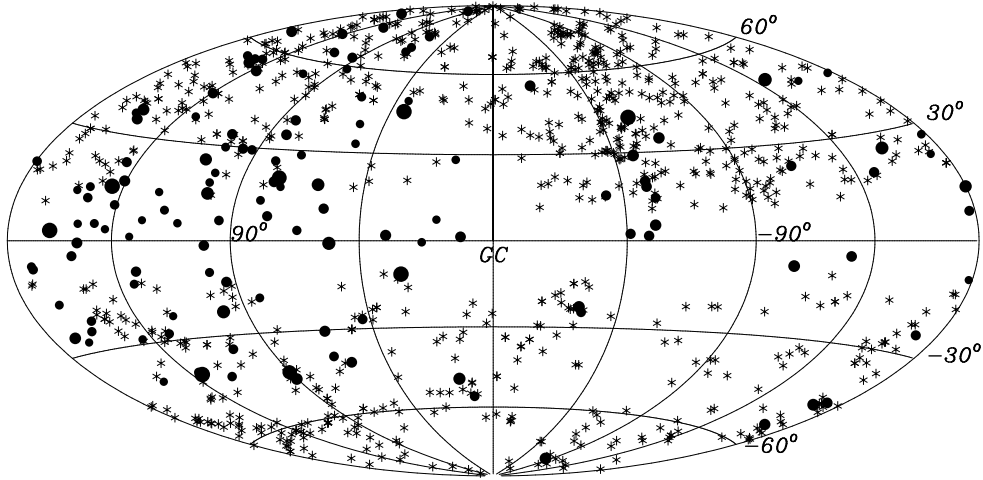} \centering
\caption{Sky map for 830 nearby VCV AGNs (*) and 135 UHECRs (dots)
in the Galactic coordinates with the deflection corrected by our GMF
toy model. Size of black dots are proportional to the cosmic-ray
energy. } \label{fig4}
\end{figure*}

\subsection{The Correlation Between UHECRs and VCV AGNs}

All sky distribution of AGNs in our selected sample and the
deflection-corrected positions of the 135 UHECR events are presented
in Figure~\ref{fig4}. We perform the correlation analysis separately
for Auger, AGASA, and Yakutsk events. The correlation results for
the UHECRs with and without deflection corrections are presented in
Table~\ref{tab4}. We also present the correlation results for
deflection corrections using the PS model \citep{ps03} and various
GMF components and their combinations in our toy model. In
Figures~\ref{fig6} and \ref{fig5}, we show the results of pair
counting and probability analysis for the deflection-corrected
UHECRs by {our} GMF toy model.

\begin{figure}
\centering
\includegraphics[width=85mm]{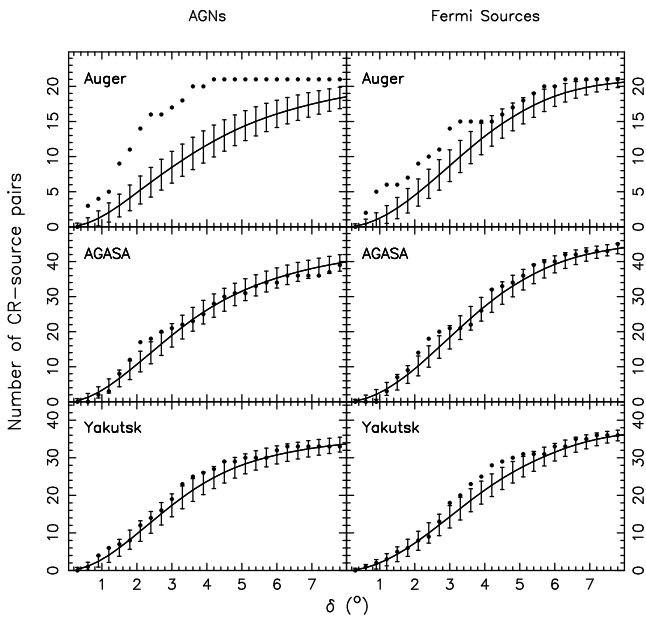} \centering
\centering
 \caption{Number of UHECR-source pairs~(dot) for nearby VCV AGNs (left
 column) and \emph{Fermi} sources (right column), as
   a function of the angular separation (bin size) $\delta$. The
   solid line is the Monte-Carlo simulated average number of
   UHECR-source pairs with errorbar of $\pm1\sigma$, derived from random
    isotropic distribution of simulated cosmic-ray events.}
\label{fig6}
\end{figure}

\begin{figure}
\centering
\includegraphics[width=0.4\textwidth]{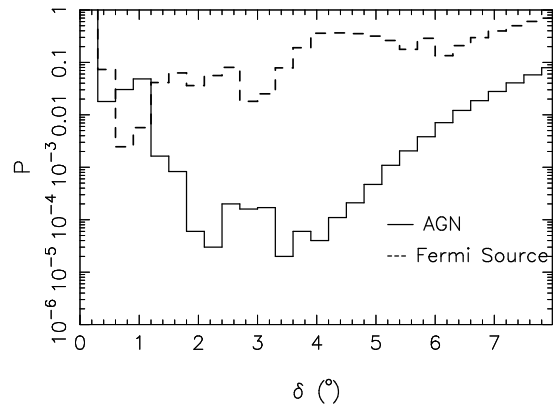} \centering
 \caption{Chance probability $P(\delta)$ as a function of the angular separation
 $\delta$ of Auger UHECR events from the VCV AGNs and the Fermi sources (dashed line), for example. }
 \label{fig5}
\end{figure}

In Figures~\ref{fig6}, \ref{fig5}, and Table 1, we found that the
number of Auger UHECR-AGN pairs is more than that of simulated
isotropic random UHECR samples, which suggests that the Auger UHECRs
are anisotropic and somehow correlated with VCV AGNs, supporting the
results of \citet{pao07,pao08}. For the deflection-corrected Auger
UHECRs by our GMF model and the PS model, the chance probability is
$2\times10^{-5}$, and the correlation significance is about
4$\sigma$, which is similar to but slightly less significant than
the results of a chance probability of $1\times10^{-6}$ and
correlation significance of 4.7$\sigma$ without deflection
correction. In {Table}~1, we found that the similar correlation
results are also presented in the cases of various GMF components
and the PS model, which indicate that some of UHECRs probably come
from a few of AGNs. The marginal correlation significance of a few
sigma ($\lesssim 5\sigma$) only suggests that most UHECRs do not
have AGN counterparts. To further evaluate the deflection effect, we
use the deflection-corrected Auger events to search for match pairs,
and compare them with the results of \citet{pao08} who did not
consider the deflections. Using the 442 nearby VCV AGNs from 12th
edition, we found 15 UHECR-AGN pairs, instead of 20 pairs reported
by \citet{pao08} with the parameters $z_{\max}=0.017$,
$\psi=3_.^{\circ}2$, and $E_{\rm th}$=57 EeV.
 Note that the fewer pairs imply the weaker correlation between
UHECRs and AGNs after deflection correction. Some of the matched
pairs reported by \citet{pao08} are probably due to random match.
The Auger UHECR-AGN correlation is also weakened when the update
Auger UHECR data are used \citep{pao09}.

As shown in Figures~\ref{fig6}, \ref{fig5}, and Table~\ref{tab4}, we
found no correlation between the AGASA and Yakutsk events and AGNs.

\begin{figure*}
\centering
\includegraphics[width=0.9\textwidth]{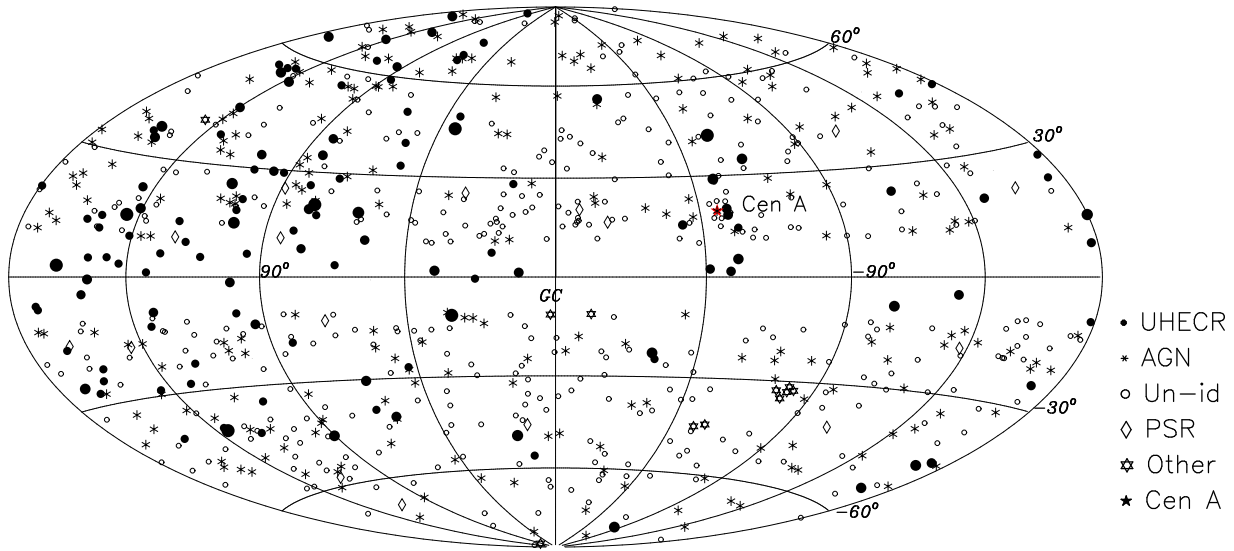}
\caption{All sky maps of 135 UHECR events and {635} $Fermi$ sources
in Galactic coordinates with the deflection corrected by {our GMF
toy} model. Black {dots} are the locations of 135 UHECRs weighted
with their energy.} \label{fig7}
\end{figure*}

\subsection{Correlation Between UHECRs and Fermi $\gamma$-ray Sources}

High energy $\gamma$-ray emissions are thought to be a distinctive
feature of the possible source of UHECRs \citep{gtt02}.
\citet{trra03} searched for possible correlation between the third
EGRET sources and AGASA UHECR events, no correlation was found. The
GeV AGNs detected by the Fermi/LAT should be most energetic AGNs
which may have powerful ability to accelerate the charged particles
to ultrahigh energy bands. Most important is that the selected GeV
AGNs in the Fermi catalog are relatively complete above the
$\gamma$-ray flux threshold. We therefore investigate the possible
correlations between $Fermi$ sources and the Auger/AGASA/Yakutsk
UHECR events considering the deflection of the GMFs. At the final
stage of this work, we noticed that \citet{mio10} did the similar
work as us and studied the possible correlation between the
$Fermi$/LAT First Source Catalog with the public Auger UHECRs. They
found that 12 of the 27 Auger UHECRs arrived within 3\d{$^\circ$}1
of Fermi sources, similar to the matches of artificially random
distribution of UHECR samples. They concluded for no correlation.
However, the possible correlation of UHECRs and $Fermi$ sources {is
needed} to be re-examined after the deflections by the GMFs and the
GZK effect are considered.

 The \emph{Fermi}/LAT 1FGL catalog contains 1451 point sources, including 820 identified objects:
 62 pulsars, 295 BL Lacs, 278 FSRQs, 120 other types of AGNs, and 65 other $\gamma$-ray
sources. We remove the $Fermi$ sources with redshift $z>0.024$ by
considering the GZK cutoff, and also neglect the sources located in
the Galactic disk with $|b|\leqslant 10^\circ$. Figure~\ref{fig7}
presents the sky distributions of the remaining 635 $Fermi$/LAT
$\gamma$-ray sources and the deflection-corrected UHECR events.

The correlation results for the \emph{Fermi} $\gamma$-ray sources
and UHECRs are listed in Table~\ref{tab4}. In the right panel of
Figure~\ref{fig6}, we also present the number of UHECR-source pairs
as a function of the bin size $\delta$. In the case of
deflection-corrected Auger events by our GMF model, a small excess
appears around $\delta\sim 0_.^\circ9$, with a chance probability of
$P\sim2.5 \times 10^{-3}$, and a significance level of $\sim
4.1\sigma$. A correlation with a chance probability of $P\sim6.0
\times 10^{-4}$, and a significance level of $\sim 3.7\sigma$ is
also found for the case with deflection correction by the PS model.
These correlations with marginal significance probably indicate that
a few $\emph{Fermi}$ sources are related to the UHECRs. However,
most of the Fermi sources are not potential sources of UHECRs. No
evidence of correlation is found in the case of the observed
positions of Auger events. In the cases of the AGASA and {Yakutsk}
data, no significant correlations are found, though many UHECRs have
a Fermi source within 3$^\circ$ from their arrival positions. In
order to evaluate the effect of deflections on the correlation
analysis and compare our results with those of \citet{mio10}, we
also try to use all 1451 sources in the 1FGL for correlation with
Auger UHECRs. We found that the number of matched UHECR-source pairs
can be reproduced by the simulated isotropic UHECR samples for the
UHECR sample without deflection correction, which is consistent with
the conclusion of no correlation by \citet{mio10}. When we exclude
the sources with $|b|\leqslant 10^\circ$ and consider the deflection
correction by our GMF model, a marginal correlation is found with a
chance probability of $P\sim9.8 \times 10^{-3}$ and a significance
level of $\sim 3.1\sigma$.

$Fermi$ sources contain several types of objects, such as pulsars,
AGNs, and unidentified sources. It is interesting to see if possible
correlations exist between some types of $Fermi$ sources and UHECRs.
The correlation analysis results for the $Fermi$ AGNs and UHECRs are
presented in Table~\ref{tab4}. The $Fermi$ AGNs are weakly
correlated with AGASA UHECRs after the deflection corrected by our
 GMF model and the PS model. Three hundred seventy-three unidentified \emph{Fermi}
sources are also weakly correlated with Auger UHECRs after the
deflection corrected by the PS model. The correlation results for
UHECRs with deflection corrections using various GMF components and
their combinations in our toy model are also presented in
Table~\ref{tab4}.

$Fermi$ 1FGL catalog has 8 objects of redshift $z\leqslant0.024$:
NGC 253 ($z=0.001$), NGC 4945 ($z=0.002$), Centaurus A (Cen A,
$z=0.002$), M87 ($z=0.004$), ESO 323-G77 ($z=0.015$), NGC 6951
($z=0.005$), NGC 1275 ($z=0.018$), and M-82 ($z=0.001$). We found
that 4 of these 8 objects have UHECR counterpart(s) within
$3_.^{\circ}1$ from them, after the deflections are corrected by our
toy model. \citet{mio10} found 2 $Fermi$ objects (NGC 4945 and Cen
A) within $3_.^{\circ}1$ from 3 of the 27 Auger UHECRs. When we
consider the deflection correction, we find one Auger UHECR event
within 3.1$^{\circ}$ from NGC 4945, Cen A and ESO 323-G77. NGC 6951
also has one Yakutsk event within $3_.^{\circ}1$.

The Cen A is the nearest FR II radio galaxy \citep{ifp98}, which has
been long proposed as a possible source of UHECRs
\citep{cag78,rcp96}. Cen A was detected at MeV to GeV energies by
the Fermi/LAT \citep{fermi09}. \citet{pao07,pao08} pointed that 4 of
the 27 events were possibly associated with Cen A \citep[e.g.][]{
msp09, kot09}. From Figure~\ref{fig3}, we found that the arrival
directions of cosmic ray in the region of Cen A are not
significantly corrected by the GMF model. Two of UHECR events in the
27 published Auger data set are very close to Cen A.
Considering heavier composition of UHECRs, \citet{piran10} suggested
that Cen A is the only active potential source of heavy nuclei
UHECRs within a few Mpc for the GZK cutoff. The heavy nuclei suffer
a larger deflection which can erase any correlation with their
source. If all detected UHECRs are produced by merely Cen A, which
come to our Galaxy and suffer different deflection via different
paths and finally arrive at the Earth from various directions, they
should show some kind of concentration around the source direction
for many lighter nuclei, depending on the detailed composition and
magnetic deflection. In Figure~\ref{fig7}, there is some indication
for such a concentration within about $20\degr$ near Cen A. While,
other UHECRs coming from other very different directions may have
other accelerating sources rather than Cen A.

NGC 4945 is identified as a Seyfert galaxy \citep{vcv06}, also known
as a starburst galaxy \citep{lt09}. \citet{lt09} identified a
non-thermal source with a jet-like morphology near the AGN of NGC
4945. ESO 323-G77 is identified as a Seyfert galaxy having strong
Fe$_II$ emission \citep{fai86}, however, no compact radio core or a
radio excess {has} been detected \citep{cor03}. NGC 6951 is known as
a LINER galaxy and a bipolar outflow which seems to be associated
with a nuclear jet has been reported by \citet{sb07}. However, it is
not clear whether the possible radio jets from NGC 4945 and NGC 6951
could be the accelerator of UHECRs.

\section{Summary {and Discussion}}

We collected 135 published UHECR events including 57 UHECRs recorded
by AGASA with energy $E>40$ EeV, {51 events observed by Yakutsk,
both} located in the northern hemisphere, and 27 events with energy
$E\geqslant 57$ EeV detected by Auger located in the southern
hemisphere. We use a new GMF toy model constrained by updated
measurements to evaluate the deflection effects on the arrival
directions of UHECRs. Considering the possible deflection correction
by our toy model and the PS model, as well as the different magnetic
field components in our model, we search for the possible
correlations of UHECRs with nearby AGNs extracted from the new
13{th} VCV AGN catalog of \citet{vcv10} and the $Fermi$/LAT First
Source Catalog of $\gamma$-ray sources. We found a correlation
between the Auger UHECR events and nearby VCV AGNs with a chance
probability of $2\times10^{-5}$, and a significance level of $\sim
4\sigma$. Using the same data as \citet{pao08}, we found fewer
UHECR-AGN pairs when deflection is considered, which implies the
weakened correlation. A marginal correlation was found between the
Auger events and the first year $Fermi$ $\gamma$-ray sources with a
significance level of $\sim 4\sigma$ if the deflection by the GMF
model is considered. Some $Fermi$ sources of nearby AGNs, NGC 4945,
ESO 323-G77, NGC 6951, and Cen A, may be related to UHECRs within
$3_.^{\circ}1$. For AGASA and Yakutsk UHECRs, no evidence of
significant correlation is found for the nearby AGNs or the $Fermi$
sources because the matched pairs can be reproduced by the simulated
random isotropic UHECR samples, though some $\gamma$-ray point
sources are coincident with the UHECR events within 2$^\circ$.

The correlations of UHECRs with some astrophysical objects suggest
that at least some of the UHECRs are protons \citep{pao07,pao08}.
However, most UHECRs seem to come from various directions and do not
associate with known astrophysical objects, which indicates that the
majority of UHECRs might suffer larger deflections in the
trajectory, due to either the unknown extragalactic magnetic fields
or the heavy nuclei component of UHECRs \citep{piran10}. The
deflection of heavy UHECRs by the GMF models is proportional to the
charge of nuclei, which leads to a very large deflection angle, for
example tens of degrees for iron, and then any correlation discussed
in this work can be diminished (Gureev \& Troisky 2010). If the
primaries of the UHECRs are heavy nuclei, instead of proton, the
identification of UHECR sources would be very difficult. Obviously,
the understanding of UHECR origin will strongly depend on our
knowledge about the strength and configuration of the Galactic and
extragalactic magnetic fields, which definitely needs more
measurements \citep{ha08}.

\begin{acknowledgements}
The authors are very grateful to the referee and Professor T. Stanev
for valuable comments and suggestions. We thank Dr. Chen Wang for
help and Dr. Zhonglue Wen, Mr. Pengfei Wang and Tao Hong for helpful
discussions. We thank Professor Chih Kang Chou, Xuyang Gao, Jun Xu,
and Hui Shi for improving our English. The Fermi $\gamma$-ray point
sources are available in the Web site
http://fermi.gsfc.nasa.gov/ssc/data. Authors are supported by the
National Natural Science Foundation of China (10803009, 10821061,
and 10833003) and the National Key Basic Science Foundation of China
(2007 CB815403).
\end{acknowledgements}


\end{document}